\documentclass[12pt]{article}
\usepackage{graphicx,bm,epsf,amsmath,amsthm,amscd,amsfonts,amssymb,amsbsy,color}
\usepackage[hyphens,spaces,obeyspaces]{url}
\usepackage[colorlinks,breaklinks,allcolors=blue]{hyperref}
\usepackage[hdvips]{hypdvips} 
\usepackage{doi}
\newcommand{\be}{\begin{equation}}
\newcommand{\ee}{\end{equation}}
\newcommand{\bea}{\begin{eqnarray}}
\newcommand{\eea}{\end{eqnarray}}
\newcommand{\nn}{\nonumber}
\newcommand{\ch}{\mbox{ch\,}}
\newcommand{\sh}{\mbox{sh\,}}

\newcommand{\e}{{\rm e}}

\begin{document}
\title{The free propagator of strongly anisotropic systems with free surfaces}
\author{M.A. Shpot\\[1.5mm]
Yukhnovskii Institute for Condensed Matter Physics\\
of the National Academy of Sciences of Ukraine,\\
79011 Lviv, Ukraine}


\date{Revised on: September 25, 2025}
\maketitle
\begin{abstract}
\noindent
A brief overview of fluctuation-induced forces in statistical systems with film geometry at the critical point and the calculation of Casimir amplitudes, which characterize these forces quantitatively, is presented.
Particular attention is paid to the special features
of strongly anisotropic $m$-axis systems at the Lifshitz point, specifically, in the case of a
\emph{``perpendicular''} orientation of surfaces with free boundary conditions.
Beyond the simplest one-loop approximation, calculations of Casimir amplitudes are impossible without knowledge of the Gaussian propagator, which corresponds to the lines of Feynman diagrams in the perturbation theory.
We present an explicit expression for such a propagator in the case of an anisotropic system confined by parallel surfaces \emph{perpendicular} to one of the anisotropy axes.
Using this propagator, we reproduce the one-loop result derived earlier in an essentially different way.
The knowledge of the propagator provides the possibility of higher-order calculations in perturbation theory, especially in the context of the epsilon expansion around the upper critical dimension.
\end{abstract}

\section{Fluctuation-induced forces in constrained systems}

Theoretical studies of spatially constrained systems are particularly important because all real physical objects have finite size and are bounded by surfaces. Moreover, such objects exhibit effects, absent in idealized infinite-volume model systems. One example is the critical Casimir effect, where fluctuation-induced forces of attraction or repulsion emerge between the boundary surfaces. This effect arises from long-range fluctuations as the temperature approaches the critical point of a bulk phase transition. The magnitude of such fluctuation-induced interactions can be measured experimentally \cite{GC99,FYP05,HHGDB08} and is quantitatively characterized by the values of the so-called Casimir amplitudes. The Casimir effect in critical systems (or thermodynamic Casimir effect) is considered in a number of reviews, \cite{Krech,KG99,BDT00,Gam09,DD23}, the most recent of which \cite{DD23} pays particular attention to experimental aspects.

The name of the effect comes from the Dutch physicist Casimir, who showed in 1948 \cite{Cas48} that parallel conducting plates placed in vacuum at distance $L$ are attracted due to the existence of zero vacuum fluctuations of the electromagnetic field.
Casimir calculated the interaction energy $\delta E(L)$ between these surfaces per unit area
and obtained the result
\begin{equation}\label{CasE}
\delta E(L)/A=\hbar c\Big(-\frac{\pi^2}{720}\Big)\,L^{-3},
\end{equation}
where $A$ is the area of the plates. Subsequently, the coefficient $-\pi^2/720$
became known as the Casimir amplitude. By differentiating $\delta E(L)/A$ with respect to $L$, Casimir derived an expression for the attractive force between the plates as
\begin{equation}
|\mathcal F|=\hbar c\,\frac{\pi^2}{240}\,L^{-4}=0.013\,L_\mu^{-4}\;{\rm dyne/ cm^2},
\end{equation}
where $L_\mu$ is the distance between surfaces measured in microns. He noted that this force could be measured experimentally, which was successfully accomplished in \cite{Lam97}.

The Casimir effect was introduced to statistical physics through the remarkable work
of Fisher and de Gennes in 1978 \cite{FdG78}.
They predicted a similar attraction between parallel surfaces immersed in a fluid
at the critical point.
Based on scaling analysis of the singular part of the free energy of such a system in three-dimensional space, these authors obtained for the interaction energy of surfaces the estimate
\begin{equation}\label{FdG}
\delta E(L)/A\sim-k_B T\,L^{-2}.
\end{equation}
This is the analog of Casimir's formula \eqref{CasE} in statistical physics with the Boltzmann constant $k_{\rm B}$ and the absolute temperature $T$.
The corresponding attractive force decreases according to the power law $\sim L^{-3}$.
In two space dimensions, a logarithmic correction appears.
According to Fisher and de Gennes \cite[p. 208]{FdG78}, $\delta E(L)/A\sim k_B T\,(\ln L)/L$.

In 1981, Symanzik \cite{Sym81} showed that an analog of the Casimir effect exists
in massless (super)renormalizable quantum field theory with $g\phi^4$ interaction in
$d=4-\varepsilon$ dimensions and film geometry $\infty^{d-1}\times L$.
He proved the finiteness of Casimir energy $\delta E(L)$ for a pair of
parallel surfaces with zero Dirichlet boundary conditions to all orders of perturbation theory, and recorded (without giving any computational details) an explicit expression for this energy to first order in the coupling constant $g$ \cite[p. 12]{Sym81}.
This result was later reproduced by different means in \cite{KD91} and \cite{SCP16}
in the context of statistical physics where the same type of the field theory is appropriate.
It has been established that in $d$-dimensional space, $\delta E(L)\sim L^{-(d-1)}$. This scaling dependence generalizes the powers of $L$ that appear in \eqref{CasE} and \eqref{FdG}, across the interval $2<d\le4$, where the corresponding Casimir force scales as $F\sim L^{-d}$.

Using the Wilson-Fisher $\varepsilon$-expansion \cite{WF72}, Krech and Dietrich determined the Casimir amplitudes for systems with $n$-component order parameters and various boundary conditions \cite{KD91,KD92}.
Their work stimulated a large number of related investigations, including further theoretical developments \cite{Krech,Krech99,BDT00,DGS06,GD08},
experimental measurements \cite{GC99,FYP05,HHGDB08}, and
numerical simulations \cite{Krech97,DK04,Vas07,Vas09,Has10}.
Long-range forces between boundaries are extremely important for nanophysics and have practical applications in designing devices of minimal size, as reviewed by French et al. \cite{French10} and more recently by Dantchev \cite{Dan24}.

\subsection{Implications of spacial anisotropy}

Until this point, we have discussed homogeneous systems obeying isotropic scale invariance. This means that in the absence of boundaries, these systems remain "self-similar" when distances  $x$ along any space direction are transformed via
$x'=\ell x$, where $\ell$ is an arbitrary scale factor.
In homogeneous systems, the correlation length $\xi$ exhibits the same power law
singularities $\sim |T/T_c-1|^{-\nu}$ in all directions as the temperature approaches the bulk critical point $T_c$.

However, systems with inhomogeneities breaking the global translational invariance,
exhibit a special kind of scaling, the anisotropic scaling.
In such systems, there are one or several special directions (called anisotropy axes or $\alpha$-axes) together with the remaining "usual" $\beta$-directions.
Along the anisotropy axes, the usual scale factor $\ell$ is transformed to a non-trivial power $\ell^\theta$ where $\theta$ is the anisotropy exponent, and the coordinates are transformed via $x_\alpha'=\ell^\theta x_\alpha$ while $x_\beta'=\ell x_\beta$.
\label{pref}
In this case, correlation lengths $\xi_{\alpha}$ and $\xi_{\beta}$ obey
the asymptotic power laws $\xi_{\alpha,\beta}\sim |T/T_c-1|^{-\nu_{\alpha,\beta}}$
with different exponents $\nu_\beta$ and $\nu_\alpha=\theta\nu_\beta$.
Systems of this kind are said to possess anisotropic scale invariance \cite{BW89,Hen02,RDS11} and are termed \emph{strongly anisotropic systems}.
Significant differences of physical properties along different axes cannot be eliminated by merely changing the values of theoretical model parameters.

In nature, there exist many physical objects with such properties.
Important examples are magnetic materials such as MnP \cite{BSOC80,SBOC81,ZSK00},
solids with structural phase transitions \cite{AM80,Aharony80},
liquid crystals \cite{CL76,APP91,Sin00}, and ferroelectrics \cite{VS92,Vys-kn,Vlox11}.
They are characterized by complex phase diagrams with a special (multi)critical
Lifshitz point \cite{HLS75,Hor80,Sel92,Die02}, in the vicinity of which
anisotropic scale invariance is valid.

These phase diagrams contain a line $T=T_c(u)$ of continuous second-order phase transitions between a high-temperature disordered phase and two ordered phases: homogeneous and periodically modulated ones. The low-temperature phases are separated by a first-order transition line. This line tangentially approaches the
$T_c(u)$ curve and joins it at its inflection point as far as the non-ordering field $u$ takes a special value $u_{LP}$. This point, where the above three different phases meet, is called the Lifshitz point \cite{HLS75}, and $T_{LP}=T_c(u_{LP})$ is the temperature at which it occurs.

The spatial constraining effects in strongly anisotropic systems remain a subject of ongoing research \cite{BDS10,DW21,DWKS21,LJ21,Zdybel21}.
The distinct physical properties in the $\alpha\,$- and $\beta\,$-directions suggest that the presence of surfaces will lead to more complex effects than in isotropic systems.
In particular, the critical Casimir effect at the Lifshitz point depends on the orientation of boundaries with respect to the anisotropy axes.
Thus, two fundamentally different variants of surface orientation have been identified \cite{BDS10}:
``parallel'' orientation, where surfaces are parallel to all anisotropy axes, and
``perpendicular'' orientation, where one of the anisotropy axes is orthogonal to the surfaces.
The fluctuation-induced forces arising in strongly anisotropic systems
obey the scaling laws \cite{BDS10}
\begin{eqnarray}\label{FAS}
{\mathcal F}_C\mathop{\propto}_{L\to \infty}
\left\{\begin{array}{ll}
\sigma^{-m/4}\Delta^{\rm BC}_{\|} L^{-\lambda_{\|}},
\vspace{2mm}\\
\sigma^{(d-m)/(4\theta)}\Delta^{\rm BC}_{\perp} L^{-\lambda_{\|}/\theta},
\end{array} \right.
\quad\mbox{where}\quad \lambda_{\|}=d-m+\theta m,
\end{eqnarray}
and $m$ is the number of anisotropy axes.
The Casimir amplitudes $\Delta^{\rm BC}_{\|}$ and $\Delta^{\rm BC}_{\perp}$ correspond to parallel and perpendicular orientations. Their values also depend on boundary conditions at the surfaces as indicated by the superscripts ``${\rm BC}$''.
The origin and nature of the scale factor $\sigma^{1/4}$ should become clear in the context of Eq. \eqref{TEN} below.

The decay exponent $\lambda_{\perp}=\lambda_{\|}/\theta$ for
perpendicular orientation is $1/\theta\approx 2$ times larger
than that for parallel orientation, and the corresponding long-range effective forces
manifest themselves to a significantly lesser degree.
The estimate $1/\theta\approx 2$ stems from the fact that
the anisotropy index $\theta$ is given by $\theta=(2-\eta_{\perp})/(4-\eta_{\|})$ \cite[p. 4]{Die02}, where
the critical correlation exponents $\eta_{\perp}$ and $\eta_{\|}$ are small.
In the framework of the $\varepsilon$-expansion with $\varepsilon=4+m/2-d$, both of them are of order $O(\varepsilon^2)$ \cite{DS00,SD01}.

\section{Thermodynamics of the critical Casimir effect in strongly anisotropic systems}

In anisotropic systems with film geometry, the dependence of Casimir forces on surface orientations manifests itself in the scaling behavior of the basic thermodynamic quantity, the \emph{residual free energy} $f_{\rm res}^{a,b}(L)$.
This important quantity is conventionally defined as a part of the free energy density per unit area (see e.g. \cite{DGS06}),
\begin{equation}\label{fdec}
f_L\equiv\lim_{A\to\infty}\frac{F}{Ak_B T}=L\,f_\infty+f_s^a+f_s^b+f_{\rm res}^{a,b}(L)\;.
\end{equation}
Here, $F=-k_B T\ln Z$ is the total free energy of a $d$-dimensional system in a slab of thickness $L$, $Z$ is the corresponding partition function, and $A$ is the area
of $(d-1)$-dimensional hypersurfaces constraining the layer.
Further, $f_\infty$ is the free energy density per unit volume of the bulk system without geometric constraints,
and $f_s^a+f_s^b$ is the total free energy density per unit area of both surfaces.
The indices $a$ and $b$ denote both the surfaces themselves and the type of boundary conditions for each of them: These boundary conditions can be different.
The last term in \eqref{fdec} is the residual free energy $f_{\rm res}^{a,b}(L)$ --- the key thermodynamic quantity in our analysis.
Beyond its usual dependence on the temperature and slab thickness $L$, in the case of strongly anisotropic systems the orientation of surfaces has essential significance \cite{BDS10}.
Thus, in dealing with strongly anisotropic systems,
we shall denote the parallel and perpendicular orientations by the index $\iota=\left\{\|,\perp\right\}$.

According to the finite-size scaling theory \cite{CardyFSS,BDS10}, the function
$f_{\rm res}^{a,b}(L)$ attains the scaling form
\begin{equation}\label{fressf}
f_{\rm res}^{a,b}(L)\approx L^{-\zeta_\iota}\,\Theta_\iota^{a,b}(L/\xi_\infty)
\end{equation}
on sufficiently large length scales.
Here, $\xi_\infty$ is the correlation length of the bulk system
and $\Theta_\iota^{a,b}(x)$ is a universal scaling function depending
on the bulk universality class of the system, its geometry, types of boundaries and their orientation with respect to the anisotropy axes.
The decay exponents $\zeta_\iota$ are also universal.
For an $m$-axial strongly anisotropic system in $d$ dimensions, they are given by (cf. \eqref{FAS} and \cite[(1.4)]{BDS10})
\begin{equation}\label{zetas}
\zeta_\|=d-m+\theta\, m-1\quad\quad\mbox{and}\quad\quad
\zeta_\perp=(d-m)/\theta+m-1\;.
\end{equation}

At the critical temperature, the bulk correlation length $\xi_\infty$ becomes infinite, and the quantity $f_{\rm res}^{a,b}(L)$ behaves as
\bea\label{fres}
f_{{\rm res}}^{a,b}(L)\approx\Theta_\iota^{a,b}(0)\,L^{-\zeta_\iota}\equiv
\Delta^{\rm BC}_\iota\,L^{-\zeta_\iota}
\eea
in the limit $L\to\infty$.
For a given medium and surfaces of certain type and orientation, the scaling function value $\Theta_\iota^{a,b}(0)$ determines the Casimir amplitude: $\Theta_\iota^{a,b}(0)\equiv\Delta_\iota^{\rm BC}$. This is the proportionality constant that already appeared in \eqref{FAS},
a universal quantity depending only on the most important gross features of the system.

The fluctuation-induced force between surfaces at the Lifshitz point is given by
\begin{equation}\label{FCAS}
\frac{{\mathcal F}_C}{k_B T_{LP}}=
-\frac{\partial}{\partial L}f_{\rm res}^{a,b}(L;T=T_{LP})=
\zeta_\iota\Delta_\iota^{\rm BC}
L^{-(\zeta_\iota+1)},\qquad\iota=\left\{\|,\perp\right\}.
\end{equation}
This is equivalent to the formula \eqref{FAS} above.
Negative values of Casimir amplitudes correspond to
attractive forces between surfaces, while positive values correspond to repulsive ones.

The relations \eqref{FAS} and \eqref{FCAS} are quite general. They are valid for
$d$-dimensional short-range interaction systems with an arbitrary number of anisotropy axes
$m$ and non-classical values of the anisotropy exponent $\theta$.
In the special case of classical Gaussian theories with $d=3$, $m=2$ and
$\theta=1/2$, equations \eqref{FAS} and \eqref{FCAS} reduce to that predicted in
\cite{ADHPP92}: ${\mathcal F}_\|\propto\sigma^{-1/2}L^{-2}$ and
${\mathcal F}_\perp\propto\sigma^{1/2}L^{-4}$.

The first non-classical expressions for Casimir amplitudes $\Delta_\iota^{\rm BC}$
for strongly anisotropic slab systems with different boundary conditions
and both variants of surface orientation were obtained in \cite{BDS10}.
The case of free Dirichlet-type boundary conditions combined with the perpendicular orientation is the most complex to treat, and we shall focus on it in what follows.

\section{Anisotropic systems with free boundary conditions}

Model systems with free boundary conditions
are attractive from a practical perspective because they
provide a more or less realistic representation of corresponding confined physical objects.
However, the absence of translational invariance, which makes a key difference from systems with periodic boundary conditions, significantly complicates a mathematical treatment of such models. The presence of anisotropy presents a further challenge.

The case of parallel surface orientation in anisotropic systems with
free boundary conditions is technically relatively simple.
Thus, it was possible to obtain in \cite{BDS10} the corresponding Casimir amplitude
to order $O(\varepsilon)$, where $\varepsilon=4+m/2-d$, and in the spherical limit $n\to\infty$ of the Stanley's $n$-vector model, where $n$ is the number of the order parameter components.

Imposing the free boundary conditions of vanishing order parameter field
on boundaries $\phi=0$ in isotropic systems implies that they undergo an ordinary transition at  $T=T_c$ \cite{Die86a}.
However, for anisotropic systems with a perpendicular surface orientation the free boundary conditions correspond to the couple of conditions
\be\label{FREE}
\phi=\partial_n\phi=0
\ee
on the boundary \cite{DSP06}. Here, $\partial_n$ denotes a derivative along the inward normal to the surface, that is, along one of the anisotropy axes.
For this geometry, in \cite[(5.86)]{BDS10} the Casimir amplitude
$\Delta^{\rm FF}_\perp$  was calculated
in the Gaussian approximation, which yielded only the leading term
of zeroth order in the possible $\varepsilon$-expansion.
The superscript ${\rm FF}$ indicates that the same free boundary conditions apply to the both boundaries of the film.
\label{KOR}

To derive that result, the Gaussian partition function $Z_G$ was written in \cite{BDS10} as a functional integral
\be\label{WDEL}
Z_G=\int\!\mathcal D\phi\;\e^{-\mathcal H_0^{\rm LP}[\phi]}
\prod_{\bm r}\prod_{j=1}^2\delta\left(\bm\phi(\bm r,z_j)\right)\,
\delta\left(\partial_z\bm\phi(\bm r,z_j)\right)
\ee
following the lines of \cite{LK91,LK92}.
The functional integration is performed over field configurations
$\bm\phi(\bm x)=\{\phi_i(\bm x),\,i=1,\ldots,n\}$ for all $\bm x\equiv(\bm r,z)$ in the infinite space $\mathbb R^d$.
The geometric constraints and necessary boundary conditions are imposed by introducing the appropriate Dirac delta functions.
These ensure that the field and its normal derivative vanish on surfaces
at $z_1=0$ and $z_2=L$.
The perpendicular configuration presently considered means that the $z$-axis
coincides with one of the $\alpha$-directions, as introduced on page \pageref{pref}.
Finally, $\mathcal H_0^{\rm LP}[\phi]$ is the translationally invariant, within $\beta$ and $\alpha$ sub-spaces, Gaussian Hamiltonian of the \emph{bulk} system directly at the Lifshitz point \cite{HLS75,DS00}:
\be\label{TEX}
\mathcal H_0^{\rm LP}[\phi]=\frac{1}{2}\int d^{d-m} x_\beta\int d^m x_\alpha
\Big(|\nabla_\beta\bm\phi|^2+\sigma_0|\Delta_\alpha\bm\phi|^2\Big).
\ee
The gradient operator $\nabla_\beta$ acts along the $\beta$-directions in
the subspace $\mathbb R_\beta^{d-m}$, and
the Laplacian operator $\Delta_\alpha\equiv\nabla_\alpha^2$ acts along the anisotropy axes
in the subspace $\mathbb R_\alpha^m$ of the full physical space of dimension $d$.

The functional integration in \eqref{WDEL} performed in \cite{BDS10}
yields the Gaussian approximation for the residual free energy
$f_{\perp,\,\rm res}^{FF}(L)=\sigma_0^{(d-m)/2}\Delta^{\rm FF}_\perp L^{-(2d-m-1)}$
in the form of a double integral, where the
Casimir amplitude $\Delta^{\rm FF}_\perp=\Delta^{\rm FF}_\perp(d,m,n)$
with $1\le m\le d$ is given by
\begin{align}\label{ELF}
\Delta^{\rm FF}_\perp=
&\frac{n}{2}K_{d-m}K_{m-1}\int_0^\infty dp\,p^{d-m-1}\int_0^\infty dq\,q^{m-2}
\nn\\&
\ln\!\Big[2\e^{-2\kappa_+}\!
\Big(\ch 2\kappa_+ +\frac{\kappa_+^2}{\kappa_-^2} \cos 2\kappa_-
-1-\frac{\kappa_+^2}{\kappa_-^2}\Big)\Big].
\end{align}
Here, $K_d$ is a usual geometric factor defined as $K_d\equiv(2\pi)^{-d}S_d$, where $S_d=2\pi^{d/2}/\Gamma(d/2)$ is the surface area of a unit sphere embedded in $\mathbb R^d$. In \eqref{ELF}, we deal with the $(d-m)$- and $(m-1)$-dimensional versions of this same constant $K_d$.
A further notation is
\be\label{KMP}
\kappa_\mp=\frac1{\sqrt 2}\sqrt{\sqrt{p^2+q^4}\mp q^2}\,,
\ee
where $p$ and $q$ are the magnitudes of momentum vectors conjugate to
$\bm x_\beta\in\mathbb R_\beta^{d-m}$ and $\bm x_\alpha\in\mathbb R_\alpha^{m-1}$.
In the present geometry, all position vectors
$\bm x_\beta$ and $\bm x_\alpha$ lie in $(d{-}1)$-dimensional hyperplanes
$\mathbb R_\beta^{d-m}\oplus\mathbb R_\alpha^{m-1}$ parallel to the surfaces.
At the endpoints of the interval $1\le m\le d$ where \eqref{ELF} holds, the Casimir amplitudes
$\Delta^{\rm FF}_\perp(d,1,n)$ and $\Delta^{\rm FF}_\perp(d,d,n)$ can be  expressed as simple integrals (see \cite[(5.87), (5.88)]{BDS10}).

We emphasize that the result \eqref{ELF} was obtained in \cite[Sec. 5.2.3]{BDS10} using an artificial procedure, without explicit knowledge of the Gaussian propagator for the system currently considered.
It is not possible to proceed in the same way to derive higher orders of perturbation theory in $\varepsilon=4+m/2-d$ for an interacting $\phi^4$ theory.
To construct higher-order Feynman diagrams, a knowledge of the free propagator of the theory is necessary. We will present this result in the next section.
Furthermore, we shall show that the Casimir amplitude \eqref{ELF} can be alternatively derived using the free propagator given in \eqref{OSP} below.

\section{The free propagator of a film with perpendicular orientation of free surfaces}
\label{polp}
At the Lifshitz point, the Gaussian theory of a system with a film geometry and perpendicular orientation of free surfaces is described by the effective Hamiltonian \eqref{TEX}, where the integration along one of the anisotropy axes, $z\equiv x_\alpha^{(m)}$, is restricted to a finite interval $[0,L]$:
\be\label{TEN}
\mathcal H_{0\perp}^{\rm LP}[\phi]=\frac{1}{2}\int d^{d-m}x_\beta\int d^{m-1}x_\alpha
\int_0^Ldz\Big(|\nabla_\beta\bm\phi|^2+\sigma_0|\Delta_\alpha\bm\phi|^2\Big).
\ee
As in\eqref{TEX}, a dimensional coefficient $\sigma_0$ appears in front of $|\Delta_\alpha\bm\phi|^2$. This parameter provides the scale $\sigma_0^{1/4}$ for "measuring" distances in the parallel $\alpha$-subspace. The classical dimension of this scale factor is $[\sigma_0^{1/4}]=[x_\beta/x_\alpha^{1/2}]$.
Its renormalized counterpart $\sigma$ (see \cite{DS00,SD01,BDS10}) appeared above in relations \eqref{FAS}. However, for notational simplicity we set $\sigma_0=1$ in what follows. If needed, the dependence on this parameter can be recovered by scaling arguments.
As before, the magnitudes of momenta conjugate to the perpendicular and parallel coordinates $\bm x_\beta$ and $\bm x_\alpha$ will be denoted by $p$ and $q$, respectively.

\bigskip
To find the mixed momentum-coordinate $pqz$-representation of the free propagator
$G_{\perp}^{\rm FF}(p,q;z_1,z_2)$ corresponding to the Hamiltonian \eqref{TEN}, we must solve the inhomogeneous fourth-order differential equation with a Dirac delta function on the right-hand side,
\be\label{EGH}
\left[\Big(\frac{d^2}{dz_1^2}-q^2\Big)^2+p^2\right]G_{\perp}^{\rm FF}(p,q;z_1,z_2 )
=\delta(z_1-z_2),
\ee
for $z_1,z_2\in[0,L]$ subject to four boundary conditions (cf. \eqref{FREE})
\be\label{EGF}
\left.G_{\perp}^{\rm FF}(p,q;z_1,z_2)\right|_{z_i\in\mathcal B_i}=
\partial_{z_i} \left. G_{\perp}^{\rm FF}(p,q;z_1,z_2)\right|_{z_i\in\mathcal B_i}=0
\quad\;\mbox{with}\quad i=1,2,
\ee
where $\mathcal B_1$ and $\mathcal B_2$ denote the boundaries at $z=0$ and $z=L$.

For compactness, we use a simplified symbolic notation, in terms of which the equation (\ref{EGH}) is written as
\be\label{EGX}
\bm L\,G(x,\xi)=\delta(x-\xi),\quad\quad\mbox{where}\quad\quad
\bm L\equiv \frac{d^4}{dx^4}-q_1\frac{d^2}{dx^2}+q_0,
\ee
and $G(x,\xi)$ is the Green's function.
The constant coefficients of the differential operator $\bm L$ are denoted by $q_1$ and $q_0$ to maintain contact with the notation in \cite{Everitt57}, which will be extensively used in what follows. For our equation \eqref{EGH}, we have
\be
q_1=2q^2=2(\kappa_+^2-\kappa_-^2)\;\quad\mbox{and}\quad\;
q_0=p^2+q^4=(\kappa_-^2+\kappa_+^2)^2,
\ee
where the combinations $\kappa_\mp=(1/\sqrt 2)\sqrt{\sqrt{p^2+q^4}\mp q^2}$\, appear, the same as in \eqref{ELF}--\eqref{KMP}.

The general theory of Green's functions can be found in many monographs on differential equations and methods of mathematical physics
(see, e.g., \cite{Ince,CourantH,Smirnov42}).
In these works, the treatment of equations of order higher than second seems to be either too abstract for practical purposes or limited to descriptions of examples involving only $\bm L$ with $q_1=q_0=0$.
On the other hand, our problem is a special case of the Sturm-Liouville problem for fourth-order differential equations. This has been solved by Everitt in 1957 \cite{Everitt57} with elegance and at the same time on a very practical level.
In what follows, we derive the explicit expression for the Gaussian propagator
$G_{\perp}^{\rm FF}(p,q;z_1,z_2)$ from (\ref{EGH}) using the scheme developed in \cite{Everitt57}.

\bigskip\noindent
The general properties of the Green's function $G(x,\xi)$ are well known \cite{Ince,CourantH,Smirnov42}:
\begin{itemize}\itemsep -1.4mm
\item
it is well-defined and continuous throughout the square $x,\xi\in[0,L]$, including its boundaries, together with its first and second derivatives;
\item
as a function of $x$, for $x\in[0,\xi[$ and $x\in]\xi,L]$ it satisfies the homogeneous equation $\bm L\,G(x,\xi)=0$ with prescribed initial conditions 
and has continuous derivatives up to fourth order;
\item
at $x=\xi$ its third derivative with respect to $x$ has a finite jump given by%
\footnote{
This can be easily seen by integrating both sides of the equation
(\ref{EGX}) from $\xi-\epsilon$ to $\xi+\epsilon$ and taking into
account that the value of the integral of the delta function in this
range of integration is unity.
By writing differential equations in (\ref{EGH}) and (\ref{EGX})
we use the sign convention opposite to that employed in, e.g.,
\cite{CourantH} and \cite{Everitt57}. This results in ``+1'' on the
right-hand side of \eqref{JUMP} (cf. \cite[p. 362]{CourantH},
\cite[p. 251, Eq. (52)]{Smirnov42}).}
\be\label{JUMP}
\lim_{\epsilon\to0}\left[G'''(\xi+\epsilon,\xi)-G'''(\xi-\epsilon,\xi)\right]=1;
\ee
\item
the function $G(x,\xi)$ is unique and symmetric with respect to interchange of its argument $x$ and parameter $\xi$.
\end{itemize}

\bigskip\noindent
The four basic steps in constructing a Green's function $G(x,\xi)$ are \cite{Ince,CourantH,Smirnov42}:
\begin{enumerate}\itemsep -1.7mm
\item
Determine four linearly independent fundamental solutions $w_i\equiv w_i(x)$, $i=1,\ldots,4$ of the homogeneous equation $\bm L\,w=0$ and construct the general solution of this equation as their linear combination
\be\label{LCOM}
w(x)=C_1 w_1+C_2 w_2+C_3 w_3+C_4 w_4.
\ee
\item
From the general integral $w(x)$, determine two special solutions $w(1|x)$ and $w(x|2)$ such that each of them satisfies the prescribed boundary conditions at only \emph{one} end of the interval under consideration. The functions $w(1|x)$ and $w(x|2)$ can be written as linear combinations
\be\label{U}
w(1|x)=c_1\phi_1(x)+c_2\phi_2(x)\quad\quad\mbox{and}\quad\quad
w(x|2)=c_3\chi_1(x)+c_4\chi_2(x).
\ee
In the case of simple boundary conditions \eqref{EGF} we have $w(1|0)=w'(1|0)=w(L|2)=w'(L|2)=0$.
\item
Determine the constants $c_i$ from three continuity conditions (for the functions themselves and their first and second derivatives) and jump in the third derivatives for any interior point $\xi\in(0,L)$,
\be\label{E}
w^{(\alpha)}(\xi|2)-w^{(\alpha)}(1|\xi)=\delta_{\alpha,3},\quad
\alpha=0,1,2,3,
\ee
where $\delta_{\alpha,\beta}$ is the Kronecker's delta symbol.
As solutions of equations \eqref{E}, the coefficients $c_i$ are functions of the parameter $\xi$.
\item
Once the set of constants $c_1(\xi),\ldots, c_4(\xi)$ is known, the Green's function $G(x,\xi)$ is uniquely determined as
\be
G(x,\xi){=}\left\{\begin{array}{c}
\!\!c_1(\xi)\phi_1(x)+c_2(\xi)\phi_2(x)\quad\mbox{for}\quad 0\le x\le\xi<L\\
\!\!c_3(\xi)\chi_1(x)+c_4(\xi)\chi_2(x)\quad\mbox{for}\quad 0< \xi\le x\le L
\end{array}\right.\!\!.
\ee
\end{enumerate}

A special role in Everitt's calculation scheme \cite{Everitt57} is played by the function $P_x(u,v)$, which appears when Green's formula is written in the form \cite[p. 255]{Ince}
\be\label{LKJ}
\int_{x_1}^{x_2}dx\left[v(x)\bm L u(x)-u(x)\bm L v(x)\right]=
\left.P_x(u,v)\right|_{x_1}^{x_2},
\ee
where $0\le x_1<x_2\le L$, and $u$ and $v$ are two arbitrary functions having continuous derivatives of at least fourth order.
This formula can be easily obtained from (\ref{EGX}) by repeated integration by parts. The function $P_x(u,v)$ is called the bilinear concomitant of the differential operator $\bm L$.
It depends on $x$, as indicated by the subscript, and $x$ is also an implicit argument of the functions $u$ and $v$.

For the operator $\bm L$ defined in (\ref{EGX}), we have
\be
P_x(u,v)=q_1 W(u,v)+W(u',v')-(uv'''-u'''v),
\ee
where $W$ is the Wronskian determinant, defined for $n$ functions $u_n(x)$ as
\be
W(u_1,\ldots,u_n)=\left|\!\!\begin{array}{ccc}
u_1(x)& \ldots& u_n(x)\\
\ldots& &\ldots\\
u_1^{(n-1)}(x)& \ldots& u_n^{(n-1)}(x)
\end{array}\!\!\right|.
\ee

Since the four functions $\phi_i(x)$ and $\chi_i(x)$ with $i=1,2$ from \eqref{U} are linearly independent, their Wronskian $W\equiv W(\phi_1,\phi_2,\chi_1,\chi_2)$ is nonzero, and it can be represented as
\be
W=P_{11}P_{22}-P_{12}P_{21},
\ee
where $P_{ij}\equiv P_x(\phi_i,\chi_j)$ are independent of $x$
(note that $P_x(\phi_1,\phi_2)=P_x(\chi_1,\chi_2)=0$).

The set of equations \eqref{E} is an inhomogeneous system of linear equations for the unknown coefficients $c_i$. The determinant of this system is $W\ne 0$, and therefore \eqref{E} uniquely determines the constants $c_i$.
Solving this system leads to the expression
\be\label{THR}
G(x,\xi)=\frac{1}{W}
\left|\!\!\begin{array}{cc}
\chi_1(\xi) & \!\!\chi_2(\xi)\\
P_{21} & \!\!P_{22}
\end{array}\!\!\right|
\phi_1(x)+
\frac{1}{W}
\left|\!\!\begin{array}{cc}
P_{11} & \!\!P_{12}\\
\chi_1(\xi) & \!\!\chi_2(\xi)
\end{array}\!\!\right|
\phi_2(x)\;\;\mbox{for}\;\; 0\le x\le\xi<L
\ee
and to an analogous formula for $G(x,\xi)$ with $0<\xi\le x\le L$, where the roles of $\phi$ and $\chi$ are interchanged.

\bigskip
In the specific case of the operator $\bm L$ from (\ref{EGX}), the homogeneous equation $\bm L\,w=0$ has the fundamental set of linearly independent solutions $w_i\equiv w_i(x)$, $i=1,\ldots,4$,
\be\label{LOM}
w_1{=}\e^{-x\kappa_+}\!\cos x\kappa_-,\;\;
w_2{=}\e^{-x\kappa_+}\!\sin x\kappa_-,\;\;
w_3{=}\e^{x\kappa_+}\!\cos x\kappa_-,\;\;
w_4{=}\e^{x\kappa_+}\!\sin x\kappa_-,
\ee
which coincides with that of \cite{DSP06} where a semi-infinite system with perpendicular surface orientation is considered. Thus, the general solution of equation $\bm L\,w=0$ is the linear combination \eqref{LCOM} of functions \eqref{LOM} with $\kappa_{\mp}$ from \eqref{KMP}.

Following the above general prescriptions, we construct two functions, $w(1|x)$ and $w(x|2)$, each satisfying only one boundary condition related to the end-points of the interval $[0,L]$.
For the coefficients of first of them, the initial conditions $w(1|0)=w'(1|0)=0$ yield the constraints
\be
C_1=-C_3 \quad\quad\mbox{and}\quad\quad(C_2+C_4)\kappa_-+(C_3-C_1)\kappa_+=0\,.
\ee
Using these relations in \eqref{LCOM} allows us to construct the required solution involving a pair of arbitrary constants and two corresponding linearly independent functions.
Thus we define $w(1|x)=c_1\phi_1(x)+c_2\phi_2(x)$ with
\be
\phi_1(x)=\sh x\kappa_+\sin x\kappa_-\quad\mbox{and}\quad
\phi_2(x)=\kappa_+\e^{-x\kappa_+}\!\sin x\kappa_-
-\kappa_-\sh x\kappa_+\cos x\kappa_-\,.
\ee
The second solution that satisfies the boundary condition $w(L|2)=w'(L|2)=0$
at $x=L$ can be defined simply as
\be
w(x|2)=c_3\phi_1(L-x)+c_4\phi_2(L-x).
\ee

The Wronskian of the four functions $\phi_1(x)$, $\phi_2(x)$, $\chi_1(x)=\phi_1(L-x)$, and $\chi_2(x)=\phi_2(L-x)$ appearing in $w(1|x)$ and $w(x|2)$ is given by
\be\label{WRO}
W(\phi_1,\phi_2,\chi_1,\chi_2)=
-2\kappa_-^2\kappa_+^2(\kappa_-^2+\kappa_+^2)\,{\rm w}(p,q),
\ee
where ${\rm w}(p,q)$ is
\be\label{wpq}
{\rm w}(p,q)=\kappa_-^2 \ch 2L\kappa_+ + \kappa_+^2 \cos 2L\kappa_-
-\kappa_-^2-\kappa_+^2.
\ee

\bigskip
This information is sufficient to derive, using \eqref{THR},
the explicit expression for the free Gaussian propagator $G_{\perp}^{\rm FF}(p,q;z,z')$ for an arbitrary number of anisotropy axes $m$. With the notations $s\equiv z+z'$ and $y\equiv |z-z'|$, it is given by
\be\label{OSP}
G_{\perp}^{\rm FF}(p,q;z,z')=
\frac{h(s,y)+h(2L-s,y)-h(2L-y,y)-h(y,y)}
{4\kappa_-\kappa_+(\kappa_-^2+\kappa_+^2)\,{\rm w}(p,q)},
\ee
where the function $h(s,y)$ is defined as
\begin{align}\label{HHH}
h(s,y)=&
\kappa_+\sin\kappa_- s\Big[(\kappa_-^2+\kappa_+^2)\ch\kappa_+y
-\kappa_-^2\ch\kappa_+(2L-s)\Big]
\nonumber\\&-
\kappa_-\sh\kappa_+(2L-s)\Big[(\kappa_-^2+\kappa_+^2)\cos\kappa_-y
-\kappa_+^2\cos\kappa_-s\Big],
\end{align}
and the function ${\rm w}(p,q)$ appears in \eqref{wpq}.
As before in \eqref{KMP}, $\kappa_\mp=(1/\sqrt 2)\sqrt{\sqrt{p^2+q^4}\mp q^2}$\,.

\bigskip
It is of interest to consider the result \eqref{OSP} in the special case of a
uniaxial anisotropic system, that is $m=1$. This brings about an essential simplification, because in the present geometry and $m=1$ the only anisotropy direction goes along the $z$-axis, and there are no other $\alpha$-directions. In momentum space, the vectors $\bm p$ are $(d-1)$-dimensional, and $q=0$. Hence, both $\kappa_-$ and $\kappa_+$ from \eqref{KMP} reduce to $\kappa\equiv\sqrt{p/2}$\,.
With this in mind, let us define the function $h_1(z+z',|z'-z|))$ via
\bea\nonumber
h_1(z+z',|z'-z|)&=&
\sin\kappa(z+z')\Big\{2\,\ch\kappa|z'-z|-\ch\kappa\left[2L-(z+z')\right]\Big\}
\\&-&
\sh \kappa(z+z')\Big\{2\cos\kappa|z'-z|-\cos\kappa\left[2L-(z+z')\right]\Big\},
\eea
which is the $m=1$ version of \eqref{HHH}. In terms of this function, the simplified $m=1$ version of the Gaussian propagator \eqref{OSP} reads
\bea\nonumber
G_{\perp\,,1}^{\rm FF}(p;z,z')=
\frac{1}{8\kappa^3}\frac{1}{\ch 2L\kappa+\cos 2L\kappa-2}
\Big[h(z+z',|z'-z|)-h(|z'-z|,|z'-z|)
\\
+h(2L-(z+z'),|z'-z|)-h(2L-|z'-z|,|z'-z|)\Big].
\eea

In the limit $L\to\infty$, the function $G_{\perp\,,1}^{\rm FF}(p;z,z')$
reduces to the known $pz$ representation of the Dirichlet-type free propagator
of a uniaxial anisotropic \emph{semi-infinite} system with a perpendicular boundary orientation at the Lifshitz point, see \cite[(35)]{DSP06},
\be
G_\perp^{\rm F}(p;z,z')= G_{\rm b}(p;|z'-z|)-G_{\rm b}(p;z'+z)-
\frac{1}{2\kappa^3}e^{-\kappa (z+z')}\sin\kappa z\,\sin\kappa z',
\ee
where the bulk free propagator $G_{\rm b}$ in the $pz$ representation is given by (cf. \cite[(22)]{DSP06})
\be\label{GB1}
G_{\rm b}(p;z;m=1)=\frac{1}{8\kappa^3}e^{-\kappa z}(\sin\kappa z+\cos\kappa z).
\ee
Reproduction of known results as special cases of our general formula \eqref{OSP} corroborates its correctness. Another strong argument in this direction is provided in the next section.

\bigskip
The free Gaussian propagator $G_{\perp}^{\rm FF}(p,q;z,z')$ from \eqref{OSP} can be used in calculations of the residual free energy of strongly anisotropic
systems in a slab with perpendicular surface orientation and free boundary conditions on both surfaces.

The result \eqref{OSP} easily generalizes to the case of a "massive" theory with a non-zero temperature-like variable ("mass") $\tau_0$, which drives the system away from the Lifshitz point. This is accomplished by a simple substitution $p^2\mapsto p^2+\tau_0$ in the definitions of $\kappa_-$ and $\kappa_+$. We shall employ this modification in a practical application of the next section.

\section{Calculation of the one-loop diagram}\label{alter}

A simple application of the free propagator $G_{\perp}^{\rm FF}(p,q;z,z')$ obtained in \eqref{OSP} is a calculation of the one-loop diagram $\,\;\includegraphics[width=9pt]{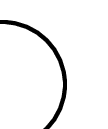}\!\!$ for the free energy of a slab.

For simplicity, let us consider the special case of a uniaxial anisotropic system with $m=1$. In this case, the single axis of anisotropy is directed along the $z$-axis perpendicular to the slab boundaries and along which the system's size is finite. All the remaining $d-1$ spatial axes parallel to the surfaces are $\beta$-directions as defined on page \pageref{pref}.

In the bulk field theory, a differentiation with respect to the mass $\tau_0$
of the Feynman integral corresponding to the ring diagram
$\;\includegraphics[width=9pt]{g0}\!$ gives an integral related to a diagram with a vertex insertion,\, $\;\raisebox{-1mm}{\includegraphics[width=9pt]{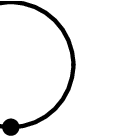}}$.
The latter is usually easier to compute.
The same holds true in the present setup of the film geometry, where the diagram
$\;\includegraphics[width=9pt]{g0}\!$ is the leading contribution to the free energy density \eqref{fdec}. The Feynman integral corresponding to the graph
$\;\raisebox{-1mm}{\includegraphics[width=9pt]{g1}}$ is then given,
in the mixed $pz$-representation, by
\begin{equation}\label{IEI}
I=\int\!\!\frac{d^{d-1}p}{(2\pi)^{d-1}}\int_0^L dz\,G_{\perp}^{\rm FF}(p;z,z).
\end{equation}
It is taken into account here that for $m=1$, the $z$-axis is an anisotropy direction.
Since there are no other anisotropy axes, the vectors $\bm q$ are absent, and only $(d-1)$-dimensional momenta $\bm p$ remain.

Since our final goal is a derivation the excess free energy following the formula \eqref{fdec},
we eliminate the bulk contribution in the integral $I$ by subtracting the bulk part of the propagator $G_{\perp}^{FF}(p;z,z)$ in its integrand. Thus we consider the integral
\begin{equation}\label{DEI}
\Delta I=\int\!\!\frac{d^{d-1}p}{(2\pi)^{d-1}}\int_0^L dz\,
\Delta G_{\perp}^{FF}(p;z,z),
\end{equation}
where $\Delta G_{\perp}^{FF}(p;z,z)=G_{\perp}^{FF}(p;z,z)-G_{\rm b}(p;z=0)$,
and $G_{\rm b}(p;z=0)$ is the $z_1\to z_2$ limit of the free bulk propagator
$G_{\rm b}(p;|z_1-z_2|)$ in the $pz$-representation at the Lifshitz point.
It is directly related to the bulk Gaussian Hamiltonian \eqref{TEX} and explicitly written down in \eqref{GB1}. We have
$G_{\rm b}(p;z=0)=1/(8\kappa^3)$ with $\kappa=\kappa_\pm|_{q=0}=\sqrt{p/2}$\,.

Thus, the integrand in \eqref{DEI} is given by
\begin{equation}
\Delta G_{\perp}^{FF}(p;z,z)=\frac{e^{-2\kappa L}}{4\kappa^3}
\frac{\hat h(z)+2-\sin 2\kappa L-\cos 2\kappa L-e^{-2\kappa L}}
{1+e^{-4\kappa L}+2e^{-2\kappa L}\cos2\kappa L-4e^{-2\kappa L}}\,,
\end{equation}
where $\hat h(z)=\hat h_1(z)+\hat h_1(L-z)$, and
\begin{equation}
\hat h_1(z)=\sin 2\kappa z\left[2-\ch2\kappa(L-z)\right]-
\sh2\kappa z\left[2-\cos 2\kappa(L-z)\right].
\end{equation}

Now we are in a position to perform the $z$ integration in \eqref{DEI}:
\begin{equation}\label{KPP}
\int_0^L dz \Delta G_{\perp}^{FF}(p;z,z)=
-\frac{1}{4\kappa^4}+\frac{L}{4\kappa^3}
\frac{e^{-2\kappa L}(2-\sin 2\kappa L-\cos 2\kappa L-e^{-2\kappa L})}
{1+e^{-4\kappa L}+2e^{-2\kappa L}\cos 2\kappa L-4e^{-2\kappa L}}\,.
\end{equation}
The first term on the right is an $L$-independent surface contribution (compare with \eqref{fdec}), while the second one is specific for the finite-size system under consideration.

At this stage, recall that both the integrals $I$ and $\Delta I$ in \eqref{IEI} and \eqref{DEI} are related to a derivative with respect to an auxiliary mass $\tau_0$ of the needed ring contribution to the free energy density \eqref{fdec}.
Therefore, restoring the mass dependence of $\kappa$ in \eqref{KPP} via
$\kappa\mapsto\kappa(\tau_0)=(1/\sqrt2)(p^2+\tau_0)^{1/4}$ (see the note at the end of the previous section),
we should do the "reverse" integration over $\tau_0$ to retrieve
the desired ring graph contribution\; $\;\includegraphics[width=9pt]{g0}\!\!$ to the free energy of the slab.
Since in the second term of \eqref{KPP}, the numerator equals the derivative of the denominator with respect to $\kappa$, such an integration over $\tau_0$ is straightforward. Thus we obtain
\bea\nonumber
&&\int\!\!\frac{d^{d-1}p}{(2\pi)^{d-1}}\int d\tau_0\int_0^L dz \Delta G_{\perp}^{FF}(p;z,z)=
-\int\!\!\frac{d^{d-1}p}{(2\pi)^{d-1}}\,\ln(p^2+\tau_0)
\\\label{KMK}&&
+\int\!\!\frac{d^{d-1}p}{(2\pi)^{d-1}}\,
\ln\left[1+e^{-4\kappa L}+2e^{-2\kappa L}\cos2\kappa L-4e^{-2\kappa L}\right].
\eea

As in \eqref{KPP}, the first, $L$-independent contribution is the surface
free energy $f_s$ from \eqref{fdec}.
The last term in \eqref{KMK} is a finite function of $L$, which agrees at
$\tau_0=0$ with the $m=1$ limit of the integral representation \eqref{ELF}.
This leads to the one-loop-approximation result
for the Casimir amplitude $\Delta_{\perp}^{\rm FF}$ at $m=1$:
\begin{equation}\label{DP1}
\Delta_{\perp}^{\rm FF}(d,1,n)=nK_{d-1}2^{-d+1}\int_0^\infty dt\,t^{2d-3}
\ln\left(1+\e^{-2t}+2\e^{-t}\cos t-4\e^{-t}\right).
\end{equation}
This result has been
derived in \cite[(5.87)]{BDS10} using a completely different method (briefly outlined
on page \pageref{KOR}) without knowledge of the free propagator $G_{\perp}^{\rm FF}(p,q;z,z')$ appearing in \eqref{OSP}.
Thus, the above calculation demonstrates how the free propagator
$G_{\perp}^{\rm FF}(p,q;z,z')$ derived in the previous section can be used in practical Feynman-graph calculations.

\section{Concluding remarks}

In this work, we have calculated the free propagator $G_{\perp}^{\rm FF}(p,q;z,z')$  for a strongly anisotropic system with film geometry, where the free surfaces are oriented perpendicular to one of the $m$ anisotropy axes. This main result is given in equation \eqref{OSP}.

Among all geometric configurations and boundary conditions considered in \cite{BDS10}, this particular case proved to be technically the most demanding.
Consequently, for this configuration the analysis of \cite{BDS10} was restricted only to the one-loop Gaussian approximation.
In the context of the epsilon expansion with $\varepsilon=4+m/2-d$, this simplest approximation gives access only to its leading $O(1)$ term.
To calculate the corresponding Casimir amplitude, a specialized procedure was employed (outlined on page \pageref{KOR}) that circumvented the need in an explicit Gaussian propagator.
In section \ref{alter}, we have reproduced this nontrivial one-loop result using an alternative approach based on the explicit expression for the propagator $G_{\perp}^{\rm FF}(p,q;z,z')$ given in \eqref{OSP}.

The calculation of $O(\varepsilon)$ and higher-order corrections remains an interesting open problem for future work. In the next order of the $\varepsilon$-expansion, computing the Casimir amplitude $\Delta_{\perp}^{\rm FF}$ requires evaluation of the two-loop diagram\, $\;\raisebox{3pt}{\includegraphics[width=10pt]{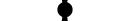}}\;$\;.
We believe that for actual calculations to order $O(\varepsilon)$, it is appropriate to use the scheme developed for homogeneous systems by Symanzik \cite{Sym81} and worked out in detail in \cite{SCP16}.
For strongly anisotropic systems with the perpendicular surface orientation considered in the present work, the overall structure of calculations and subtractions of divergent terms should remain unchanged.
At the same time, the corresponding integrands involving the propagator \eqref{OSP} will become essentially more complicated, which will eventually necessitate numerical computations.

\section*{Acknowledgements}
Part of this work was carried out at the Fakult\"at f\"ur Physik of the Universit\"at Duisburg-Essen in 2009. The author is grateful to H.W. Diehl for suggesting the problem, stimulating discussions, warm hospitality, and financial support. The author acknowledges the helpful assistance of Claude Sonnet 4 by Anthropic in translation from Ukrainian and of Gemini by Google in final editing. Financial support through the research project ``Analytical and numerical approaches to analysis of cooperative behavior in complex systems'' by the National Academy of Sciences of Ukraine under state registration number 0123U100238 is gratefully acknowledged.

The author is deeply grateful to all warriors of the Ukrainian Armed Forces, living and fallen, for making it possible to continue this research work.

\providecommand{\href}[2]{#2}
\addcontentsline{toc}{section}{References}

\end{document}